# Welding of Carbon Nanotubes to solid Surfaces using Microwave-Plasma

*Pejman Talemi*[1], George P. Simon[2]*

1: School of Chemical Engineering, University of Adelaide, Adelaide SA 5005, Australia

2: Department of Materials Engineering, Monash University, Clayton, Vic 3800, Australia

* Corresponding author. Tel: +61-8-8314646. Email:
pejman.talemi@adelaide.edu.au

**Introduction**
Because of their unique properties of carbon nanotubes such as good electrical and thermal conductivity[1-3], high mechanical strength[4, 5], and low threshold voltage for field emission[6-8], they became the focus of much attention in different fields of science and technology. For some applications such as electron field emission cathodes, only nanotubes on the surface contribute to the observed properties. Therefore many research groups are attracted to develop new techniques for engineering of carbon nanotubes on the surface. Most of these researches are focused on formation of a film or thin layer of nanotubes on the surface[9-12]. However, a few groups have tried to find methods for fixing nanotubes on the surface by welding them to the surface using microwave irradiation[13, 14].

Microwave irradiation is the electromagnetic radiation with frequencies ranging from 300MHZ to 300GHZ. By coupling this electromagnetic energy with material through molecular interaction, microwave energy can be used for heating of materials[15]. Due to their high conductivity carbon nanotubes are excellent microwave absorbing materials. It is reported that upon absorption of microwave energy by nanotubes the temperature can rise to close to 2000 °C[16]. Zhang et al[17] have suggested that this microwave heating process could be used to make polymer composites from stacks of alternating layers of polymer sheets and nanotube layers. In the next step Wang et al has reported the use of microwave irradiation for welding of carbon nanotubes to polymeric surfaces as a method for bonding of two polymer sheets[18] or making flexible electron emitters[7]. However, bonding of nanotubes to thermoplastic polymeric surfaces can easily be performed by dispersing carbon nanotubes in an appropriate solvent and using doctor blading or dip coating to apply it to polymeric surfaces. As shown in Figure 1, a dispersion of carbon nanotubes in acetone and doctor blading it on a polycarbonate film can be used for bonding nanotubes to a thermoplastic surface.

Considering that carbon nanotubes are excellent microwave absorbers, microwave irradiation has been previously used for oxidation and modification of CNTs. However long exposure of nanotubes to microwave irradiation can lead to damaging their structure[16,19]. In fact microwave irradiation of nanotubes in air results in fast oxidation, ignition and burning of nanotubes in a few seconds. Consequently using microwave for welding nanotubes to polymeric surfaces may result in combustion of the polymer substrate as well.

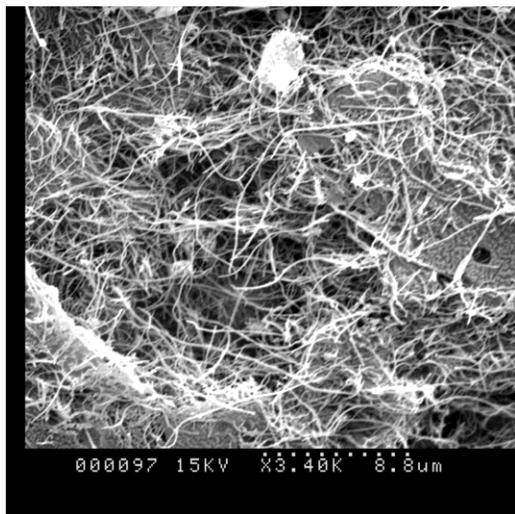

**Fig 1.** Welding of carbon nanotubes to a polycarbonate surface using a dispersion of nanotubes in acetone.

In our previous works we have shown that an unmodified kitchen microwave oven can be used for formation of plasma inside a quartz tube under vacuum. In this method, the absence of oxygen avoids sever oxidation and combustion of samples. In addition plasma is a good microwave absorber and can protect carbon nanotubes to some extent[20]. We have used this method for modification of nanotubes[20, 21], preparation of graphene sheets[22], carbon nanofibers[23, 24] and carbon nanospheres[23, 24].

In this study we will report a how this microwave-plasma method can be utilized for welding of carbon nanotubes to different surfaces, such as polystyrene as a thermoplastic polymer, lime-soda glass and borosilicate glass.

**Experimental**

Forests of vertically-aligned MWCNTs were produced via chemical vapor deposition in quartz CVD reactor tube on a catalytic bed of iron with a flow through of helium and acetylene at CSIRO and supplied for this work. The MWCNTs have an average diameter of 50nm and average length of up to 500μm. CNTs were dispersed in ethanol by mild sonication in a ultrasonic bath. The desired substrate was coated with the dispersion and dried in the oven at 70 ºC this process repeated for several time till a film with a thickness of about 0.3 mm was formed on the surface. In the next step sample was placed in a quartz vacuum tube. A rotary pump was used to reduce the pressure of the tube to 50-100 mbar. The tube was then placed in an unmodified kitchen microwave oven (NEC N920E Kitchen) and microwave irradiation with different durations of time ( 3 s for polystyrene, 60 s for lime-soda glass and 120 s for borosilicate glass) was applied to the samples. The plasma in this process is readily observed as a result of its emission of a bright, colorful light. Long exposure times may result in temperatures as high as 1000°C[22] and melt the substrate. After cooling down the samples to room temperature ultrasonic cleaning in ethanol was used for 20 min to remove the loosely or non-bonded nanotubes, finally after drying of samples were further cleaned by rubbing with a paper tissue (as a mechanical abrasive method).

**Result and discussion**

In order to avoid damaging the substrates due to the high temperatures of the microwave-plasma process, the process time was estimated using the glass transition temperature of the substrates and the temperature of the process. In order to measure the temperature of the process a blank run without any sample was performed and the temperature of the tube was measured using a Raytek Thermalert 30 infrared temperature sensor. Figure 2 demonstrates the measured temperature of the process as a function of time.
Considering the typical glass transition temperature of substrates used in this process (polystyrene ~100ºC, soda-lime glass ~570ºC, borosilicate glass ~820ºC) a process time of 3, 60 and 120 seconds has been used for these substrates, respectively.

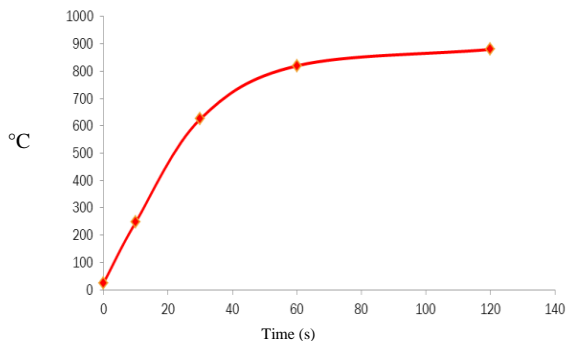

**Fig 2.** Rise of temperature as a function of microwave-plasma process time

SEM images (Figure 3) of the surface shows that CNTs are well welded to the surface through a "melt-and-glue" mechanism that remains attached to the surface after such vigorous attempts for removing them. As expected, because of lower melting point of polystyrene, welding of carbon nanotubes on polystyrene surface can be achieved with better efficiency, but higher softening point (and higher viscosity after softening point) prevents nanotubes from forming a continuous layer on glass surfaces.

Raman spectroscopy was employed to evaluate the extent of damage to graphitic structure of nanotubes in this process. Considering the Raman background of polystyrene and soda lime glass, this experiment was only conducted on borosilicate samples. In addition the borosilicate sample had the longest exposure time to microwave plasma, thus can provide information about the maximum possible damage to nanotubes. There are two independent peaks between 1000 cm$^{-1}$ and 2000 cm$^{-1}$ in the Raman spectrum of carbon nanotubes. The band at about 1580 cm$^{-1}$, labelled as the G-band, that corresponds to an E$_{2g}$ mode of graphite and is related to the vibration of sp$^2$-bonded carbon atoms in a two dimensional hexagonal lattice such as in a graphite layer, while the band at 1310 cm$^{-1}$, called the D band represents the dispersive, defect-induced band[25]. Due to this quantitative relationship between the intensity of these peaks and the amount of defects or amorphous carbon, monitoring the change in the peak intensity ratio of G and D peaks ($I_G/I_D$) is an accepted method to examine formation of defects in the nanotube structure [26-30]. Increasing the amount of defects or amorphous carbon should cause $I_G/I_D$ to decrease. $I_G/I_D$ ratio for the pristine carbon nanotube samples was found to be 0.71. Exposing nanotubes to 120 seconds of microwave-plasma environment resulted in decreasing this value to 0.64. This indicates that some defects are being introduced to the graphitic structure of the nanotubes in this process.

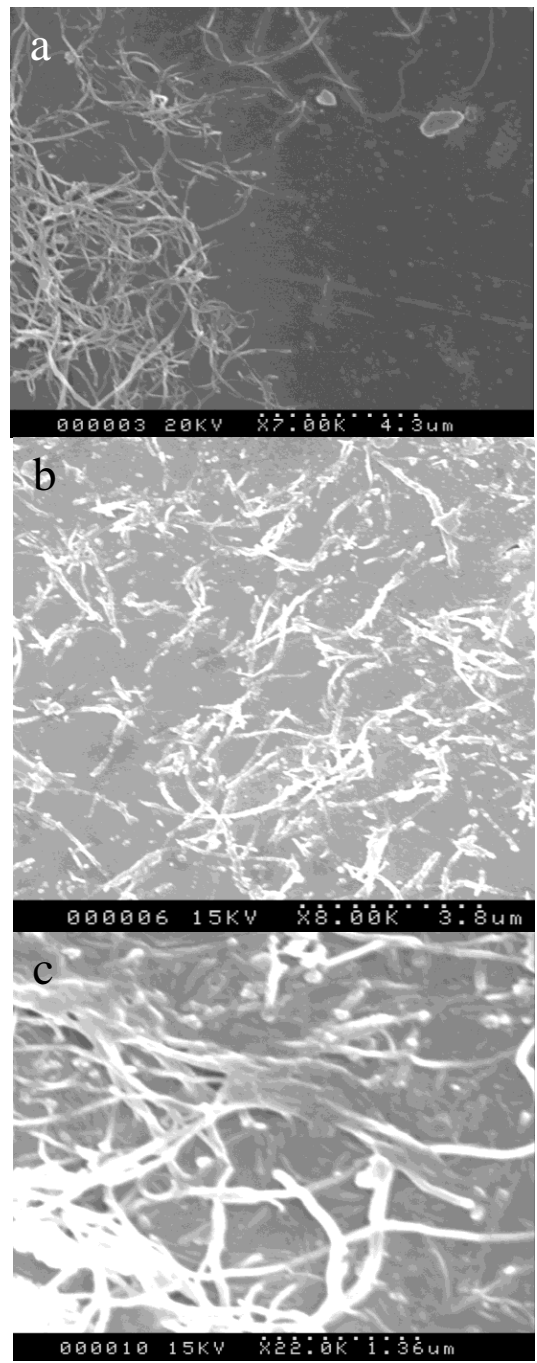

**Fig. 3** CNTs welded on a) polystyrene, b) soda-lime glass and c) borosilicate glass substrates using MW-plasma.

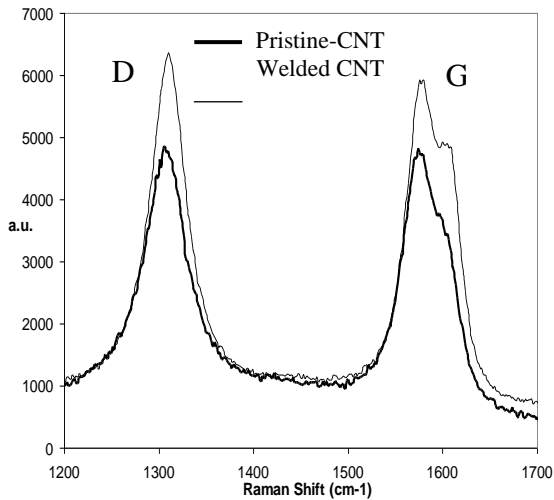

**Fig. 4** Raman Spectra of CNT1 and CNTP60 sample as an example showing changes in the relative intensity of D and G Peaks

Cancado et al [31] tried to find a relationship between the nano-graphite crystallite lateral size (L$_a$) and the ratio of D and G peaks.

$$L_a(nm) = (2.4\times10^{-10})\lambda_l^4 (\frac{I_G}{I_D})$$

where $\lambda_l$ is the wavelength of excitation laser (in nanometers) used in Raman experiments. Using the L$_a$ value, instead of the more common $I_G/I_D$ ratio, and by taking the effect of the laser excitation energy into account, the comparison of the data from different papers and reports (where different lasers are used) becomes possible. It should be noted that this equation is derived from perfect, graphitic samples, and that the intensity of the D band in the samples described here arises not just from the edges, but also from point defects, functional groups, Stone-Wales and other defects. Calculated La values for samples based on $I_G/I_D$ ratio and the wavelength of the excitation laser (782 nm) show that the value of La for the pristine nanotubes was 63.7nm, while after welding process his value has decreased to 57.4nm. This decrease in $I_G/I_D$ (and consequently L$_a$) must be due to the introduction of new structural defect sites, or minor damages such as the cutting or unzipping of nanotubes[21, 32]. Introducing some defects to carbon nanotubes is beneficial for some applications such as field emission cathodes[21]. However, comparing this change to the changes caused by other common chemical treatment methods such as acid treatment or thermal treatment[20], shows that in fact this method has not had a significant effect of the structure of carbon nanotubes.

## Conclusion

The work here has demonstrated that microwave welding of carbon nanotubes onto plastics and glasses can be achieved using a simple microwave-plasma method. Due to the lower softening point and viscosity of polystyrene (compared to glass samples), welding of carbon nanotubes on polystyrene surface can potentially be achieved with higher efficiency, whilst the high softening point (and melt viscosity) of lime-soda and borosilicate glasses prevents the nanotubes from forming a continuous layer. Raman results confirm that some defects are introduced to the CNT graphitic structure. However, these defects are not significant and can actually be useful for some application.

**References**

1. Du, F.; Fischer, J. E.; Winey, K. I., Effect of nanotube alignment on percolation conductivity in carbon nanotube/polymer composites. *Physical Review B (Condensed Matter and Materials Physics)* **2005,** 72, (12), 121404-4.
2. Astorga, H. R.; Mendoza, D., Electrical conductivity of multiwall carbon nanotubes thin films. *Optical Materials* **2005,** 27, (7), 1228-1230.
3. Haggenmueller, R.; Guthy, C.; Lukes, J. R.; Fischer, J. E.; Winey, K. I., Single Wall Carbon Nanotube/Polyethylene Nanocomposites: Thermal and Electrical Conductivity. *Macromolecules* **2007,** 40, (7), 2417-2421.
4. T. Ramanathan; H. Liu; Brinson, L. C., Functionalized SWNT/polymer nanocomposites for dramatic property improvement. *Journal of Polymer Science Part B: Polymer Physics* **2005,** 43, (17), 2269-2279.
5. McCullen, S. D.; Stevens, D. R.; Roberts, W. A.; Ojha, S. S.; Clarke, L. I.; Gorga, R. E., Morphological, Electrical, and Mechanical Characterization of Electrospun Nanofiber Mats Containing Multiwalled Carbon Nanotubes. *Macromolecules* **2007,** 40, (4), 997-1003.
6. Umnov, A. G.; Matsushita, T.; Endo, M.; Takeuchi, Y., Field emission from flexible arrays of carbon nanotubes. *Chemical Physics Letters* **2002,** 356, (3-4), 391-397.
7. Wang, C. Y.; Chen, T. H.; Chang, S. C.; Chin, T. S.; Cheng, S. Y., Flexible field emitter made of carbon nanotubes microwave welded onto polymer substrates. *Applied Physics Letters* **2007,** 90, (10), 103111-3.
8. Hojati-Talemi, P.; Hawkins, S. C.; Huynh, C. P.; Simon, G. P., Highly efficient low voltage electron emission from directly spinnable carbon nanotube webs. *Carbon* **2013,** 57, 169-173.
9. Pulickel, M. A., Aligned carbon nanotubes in a thin polymer film. *Advanced Materials* **1995,** 7, (5), 489-491.
10. H. Shimoda, S. J. O. H. Z. G. R. J. W. X. B. Z. L. E. M. O. Z., Self-Assembly of Carbon Nanotubes. *Advanced Materials* **2002,** 14, (12), 899-901.
11. Xue, W.; Cui, T., Characterization of layer-by-layer self-assembled carbon nanotube multilayer thin films. *Nanotechnology* **2007,** 18, (14), 145709.
12. L. Ci, S. M. M. X. L. R. V. P. M. A., Ultrathick Freestanding Aligned Carbon Nanotube Films. *Advanced Materials* **2007,** 19, (20), 3300-3303.
13. Asaka, K.; Nakahara, H.; Saito, Y., Nanowelding of a multiwalled carbon nanotube to metal surface and its electron field emission properties. *Applied Physics Letters* **2008,** 92, (2), 023114-3.

14. Xu, Z. W.; Zhao, Q. L.; Sun, T.; Guo, L. Q.; Wang, R.; Dong, S., Welding method for fabricating carbon nanotube probe. *Journal of Materials Processing Technology* **2007,** 190, (1-3), 397-401.
15. Collin, R. E., *Foundations for Microwave Engineering*. McGraw-Hill: New York, 1966.
16. Imholt, T. J.; Dyke, C. A.; Hasslacher, B.; Perez, J. M.; Price, D. W.; Roberts, J. A.; Scott, J. B.; Wadhawan, A.; Ye, Z.; Tour, J. M., Nanotubes in Microwave Fields: Light Emission, Intense Heat, Outgassing, and Reconstruction. *Chem. Mater.* **2003,** 15, (21), 3969-3970.
17. Zhang, M.; Fang, S.; Zakhidov, A. A.; Lee, S. B.; Aliev, A. E.; Williams, C. D.; Atkinson, K. R.; Baughman, R. H., Strong, Transparent, Multifunctional, Carbon Nanotube Sheets. *Science* **2005,** 309, (5738), 1215-1219.
18. Wang, C. Y.; Chen, T. H.; Chang, S. C.; Cheng, S. Y.; Chin, T. S., Strong Carbon-Nanotube-Polymer Bonding by Microwave Irradiation. *Advanced Functional Materials* **2007,** 17, (12), 1979-1983.
19. Vazquez, E.; Georgakilas, V.; Prato, M., Microwave-assisted purification of HIPCO carbon nanotubes. *Chemical Communications* **2002**, (20), 2308-2309.
20. Hojati-Talemi, P.; Cervini, R.; Simon, G., Effect of different microwave-based treatments on multi-walled carbon nanotubes. *Journal of Nanoparticle Research* **2010,** 12, (2), 393-403.
21. Hojati-Talemi, P.; Simon, G. P., Enhancement of field emission of carbon nanotubes using a simple microwave plasma method. *Carbon* **2011,** 49, (2), 484-486.
22. Hojati-Talemi, P.; Simon, G. P., Preparation of graphene nanowalls by a simple microwave-based method. *Carbon* **2010,** 48, (14), 3993-4000.
23. Hojati-Talemi, P.; Azadmanjiri, J.; Simon, G. P., A simple microwave-based method for preparation of Fe3O4/carbon composite nanoparticles. *Materials Letters* **2010,** 64, (15), 1684-1687.
24. Hojati-Talemi, P.; Asghari-Khiavi, M.; Simon, G., Preparation of carbon nanoparticles and nanofibers by a simple microwave based method and studying the field emission properties. *Materials Chemistry and Physics* **2011,** 127, (1–2), 156-161.
25. Yu, G.; Gong, J.; Wang, S.; Zhu, D.; He, S.; Zhu, Z., Etching effects of ethanol on multi-walled carbon nanotubes. *Carbon* **2006,** 44, (7), 1218-1224.
26. Antunes, E. F.; Lobo, A. O.; Corat, E. J.; Trava-Airoldi, V. J., Influence of diameter in the Raman spectra of aligned multi-walled carbon nanotubes. *Carbon* **2007,** 45, (5), 913-921.
27. Basca, W. S.; Ugrate, D.; WChatelain, A.; De Heer, W. A., High-Resolution electron microscopy and inelastic light scattering of purified multishelled carbon nanotubes *phys. rev.* **1994,** B50, 15473.
28. Rosca, I. D.; Watari, F.; Uo, M.; Akasaka, T., Oxidation of multiwalled carbon nanotubes by nitric acid. *Carbon* **2005,** 43, (15), 3124-3131.
29. Osswald, S.; Gogotsi, M. H. Y., Monitoring oxidation of multiwalled carbon nanotubes by Raman spectroscopy. *Journal of Raman Spectroscopy* **2007,** 38, (6), 728-736.
30. Li, W.; Bai, Y.; Zhang, Y.; Sun, M.; Cheng, R.; Xu, X.; Chen, Y.; Mo, Y., Effect of hydroxyl radical on the

structure of multi-walled carbon nanotubes. *Synthetic Metals* **2005,** 155, (3), 509-515.

31. Cancado, L. G.; Takai, K.; Enoki, T.; Endo, M.; Kim, Y. A.; Mizusaki, H.; Jorio, A.; Coelho, L. N.; Magalhaes-Paniago, R.; Pimenta, M. A., General equation for the determination of the crystallite size L[sub a] of nanographite by Raman spectroscopy. *Applied Physics Letters* **2006,** 88, (16), 163106-3.

32. Kosynkin, D. V.; Higginbotham, A. L.; Sinitskii, A.; Lomeda, J. R.; Dimiev, A.; Price, B. K.; Tour, J. M., Longitudinal unzipping of carbon nanotubes to form graphene nanoribbons. *Nature* **2009,** 458, (7240), 872-876.